# Virtual gravitational dipoles: The key for the understanding of the Universe?

Dragan Slavkov Hajdukovic[a, b]
[a]Physics Department, CERN; CH-1211 Geneva 23
[b]Institute of Physics, Astrophysics and Cosmology; Cetinje, Montenegro
E-mail: dragan.hajdukovic@cern.ch

**Abstract.** Before the end of this decade, three competing experiments (ALPHA, AEGIS and GBAR) will discover if atoms of antihydrogen fall up or down. We wonder what the major changes in astrophysics and cosmology would be if it is experimentally confirmed that antimatter falls upwards. The key point is: If antiparticles have negative gravitational charge, the quantum vacuum, well established in the Standard Model of Particles and Fields, contains virtual gravitational dipoles. The main conclusions are: (1) the physical vacuum enriched with gravitational dipoles is compatible with a cyclic universe *alternatively* dominated by matter and antimatter, without initial singularity and without need for cosmic inflation; (2) the virtual dipoles might explain the phenomena usually attributed to dark matter and dark energy. While what we have presented is still far from a complete theory, hopefully it can stimulate a radically different and potentially important way of thinking.

## 1. Introduction

Recently the ALPHA Collaboration has performed an important proof-of-principle experiment [1] that yields directly measured limits on the ratio of the *gravitational charge* to inertial mass of atoms of antihydrogen. The achievement of ALPHA Collaboration and successful preparation of two other experiments, AEGIS [2] and GBAR [3], give us certainty that the gravitational charge of antihydrogen will be measured before the end of this decade.

The present Article is a reflection on radically new astrophysics and cosmology that must be developed *if antiparticles have a negative gravitational charge*.

The first thought is that the gravitational properties of antimatter can have a major impact, *only if* the Universe contains *comparable* quantities of matter and antimatter. Therefore, in our Universe which is apparently dominated by matter, the eventual discovery of negative gravitational charges will not force us into major changes in astrophysics and cosmology. Following this line of thinking, the proponents of the gravitational repulsion between matter and antimatter have proposed alternative cosmology [4, 5] based on the assumption of equal quantities of matter and antimatter in the Universe, with antimatter hidden in cosmic voids.

However, thanks to the existence of the quantum vacuum, well established [6-8] in the Standard Model of Particles and Fields, the gravitational properties of antimatter can play a crucial role in a Universe *dominated* by matter, without any need for hidden antimatter.

Before the foundation of Quantum Field Theory, the physical vacuum was a synonym for nothing. However in quantum field theory "nothing's plenty", as nicely said by Aitchison in his classical review [6] for non-specialists readership. More precisely, the physical (or quantum) vacuum is *the ground state* (a state of minimum energy) of the considered system of fundamental fields. The other states of the system are 'excited' states, containing quanta of excitation, i.e. particles. There are no particles in the vacuum (in that sense the vacuum is empty); but the vacuum is plenty of short-living virtual particle-antiparticle pairs which in permanence appear and disappear (as is allowed by time-energy uncertainty relation $\Delta E \Delta t \geq \hbar/2$ ).

Quantum vacuum should be considered as a *new state* of matter-energy, completely different from familiar states (gas, liquid, solid, plasma…) but *as real as* they are [6-8]. Popularly speaking, quantum vacuum is an "ocean" of short living, virtual particle-antiparticle pairs (like quark-antiquark, neutrino-antineutrino and electron-positron pairs). According to our best knowledge: (1) quantum vacuum is a state with perfect *symmetry* between matter and antimatter; a particle *always* appears in pair with its antiparticle, which is totally different from mysterious matter-antimatter *asymmetry*, i.e. the fact that everything on the Earth (and apparently in the Universe) is made only from matter, with only traces of antimatter; (2) contrary to all other states of matter-energy which are composed from the long living particles (electrons and protons in stars and flowers, have existed before them and will exist after them), the quantum vacuum is a state with extremely *short living* virtual particles and antiparticles (for instance, the lifetime of a virtual electron-positron pair is only about $10^{-22}$ seconds).

While the existence of the quantum vacuum is an inherent part of the Standard Model of Particles and Fields, it is systematically neglected in Astrophysics and Cosmology; not because we are unaware of the possible gravitational impact of the quantum vacuum, but because no one knows its gravitational properties. The attempt to interpret dark energy as vacuum energy was brutally halted by the cosmological constant problem [9, 10]; theoretically predicted dark energy density is many orders of magnitude larger than the observed one.

Let us briefly consider two important phenomena in quantum electrodynamics, which are important for the understanding of the consequences of the hypothetical negative gravitational charge of antiparticles.

The first illuminating phenomenon coming from quantum electrodynamics is known as Schwinger's mechanism [11, 12]. A virtual electron-positron pair might be *converted* into a real one by a sufficiently strong external electric field which accelerates electrons and positrons in *opposite* directions. For a constant acceleration $a$ (which corresponds to a constant electric field), the particle creation rate per unit volume and time, can be written as:

$$\frac{dN_{m\bar{m}}}{dtdV} = \frac{c}{\lambdabar_m^4} \left(\frac{a}{a_{cr}}\right)^2 \sum_{n=1}^{\infty} \frac{1}{n^2} \exp\left(-n\frac{a_{cr}}{a}\right), \quad a_{cr} \equiv \pi \frac{c^2}{\lambdabar_m} \tag{1}$$

which is the famous Schwinger formula [11,12], with $\lambdabar_m \equiv \hbar/cm$ being the reduced Compton wavelength of a particle with mass $m$. In simple words, a virtual pair can be converted to a real one (i.e. real particle-antiparticle pairs can be created from the quantum vacuum!), by an external field which, during their short lifetime, can separate particle and antiparticle to a distance of about one reduced Compton wavelength. It is important to understand, the Schwinger mechanism is valid *only* for an external field that has tendency to *separate* particles and antiparticles. Hence, equation (1) can be used for the gravitational field, *only if*, particles and antiparticles have *the gravitational charge of the opposite sign*. As we will argue, the gravitational version of Schwinger's mechanism *excludes* the possibility of the gravitational collapse of Universe to a singularity. Instead, at a macroscopic size (larger than the size after cosmic inflation in the Standard Cosmology) matter of our Universe would be converted to antimatter. Hence, it is possible that our Universe (or better to say our cycle of the Universe) was born with a macroscopic size, without initial singularity and cosmic inflation, providing a simple explanation of the matter-antimatter asymmetry: We live in a Universe dominated by matter because the previous one was dominated by antimatter.

The second significant fact in quantum electrodynamics is: Virtual pairs of charged particles (for instance, electron-positron or quark-antiquark pairs) *behave as virtual electric dipoles*. Consequently, in an external electric field, the polarization of the quantum vacuum, analogous to the familiar polarization of a dielectric should be expected. In particular, the vacuum around an electron might be polarized. An electron attracts virtual particles with a charge of the opposite sign; hence there is a vacuum *screening* effect around the electron. What we measure at large distances is "*screened*" charge and it must be less than the "bare" charge. Thus, the observed charge of an electron $e$, or equivalently the fine structure constant ($\alpha \equiv e^2/4\pi\varepsilon_0 \hbar c$) should be position dependent. Today, this theoretical prediction is a confirmed reality; for instance, at large distances $\alpha^{-1} = 137.036$, while at the shortest distances probed so far [7] $\alpha^{-1} = 128.886$ (i.e. because of the quantum vacuum the electric charge of electron is about 4% greater!), which is in perfect agreement with theoretical calculations. If antimatter falls up (i.e. if particles and antiparticles have gravitational charge of the opposite sign), virtual pairs in the physical vacuum are also gravitational dipoles and it can be argued that the quantum vacuum enriched with gravitational dipoles has potential to explain phenomena usually attributed to the hypothetical dark matter and dark energy.

In brief, a surprising outcome of three experiments (ALPHA, AEGIS and GBAR) would have the importance of a new scientific revolution. The crucial point of the current model of the Universe is that the *content of the Universe* (baryonic matter + *dark matter* + *dark energy*) *is immersed in the classical* (non-quantum) vacuum. Instead, we might be forced to consider a Universe which is the inseparable union of

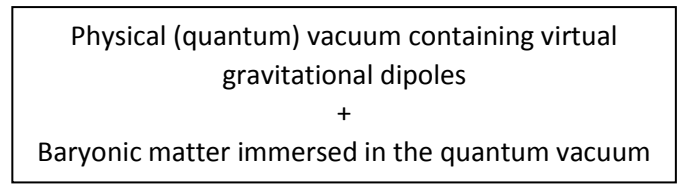

Physical (quantum) vacuum containing virtual gravitational dipoles
+
Baryonic matter immersed in the quantum vacuum

From the point of view of contemporary physics it is imperative [9, 10] to find how to include the quantum vacuum in cosmological models and, of course, it would be a great advantage if baryonic matter is the only content of the Universe. The existence of virtual gravitational dipoles might be cornerstone for such an achievement.

## 2. Cyclic Universe without initial singularity and cosmic inflation

According to astronomical observations we live in an expanding Universe. Hence, the size of the Universe was smaller in the past. How much smaller? Smaller than our Galaxy, smaller than our Solar System, smaller than an electron, or even smaller than the Planck length, as conjectured in the Standard Big Bang Cosmology? The study of the cosmic microwave background reveals that the linear size (or more precisely the cosmic scale factor $R(t)$ in the Friedmann–Lemaître–Robertson–Walker metric) was more than a thousand times smaller than today; hence, a Universe smaller than our Galaxy is presumably a fact, but everything before it is just a speculation. In the framework of contemporary physics there is no known mechanism to stop the gravitational collapse; hence, our imagined trip backward in time must end with a singularity and not at a macroscopic size. The initial singularity is one of the inherent problems of Standard Cosmology [13, 14]and one of reasons to invoke cosmic inflation [15] i.e. an expansion of the early Universe (within the first $10^{-30}$ seconds), with a speed more than twenty orders of magnitude greater than the speed of light. However, as we will show below, if there is gravitational repulsion between matter and antimatter, there is a physical mechanism to prevent gravitational collapse to singularity and to eliminate the need for cosmic inflation.

Equation (1) contains a sum of exponential functions with negative exponents; hence, the particle creation rate is significant only for a gravitational field $a$ greater than the critical value $a_{cr}(m) = \pi c^2/\lambdabar_m$. Let us compare the critical acceleration $a_{cr}(m)$ with the gravitational acceleration $g_S = GM/R_S^2 \equiv c^2/2R_S$ at the Schwarzschild radius ($R_S = 2GM/c^2$) of a black hole with mass $M$; the comparison leads to the conclusion $a_{cr} >> g_S$, i.e. a virtual pair can be converted to a real one *only deep inside the horizon* of a black hole.

Now, the qualitative picture of the expected phenomena is very simple and beautiful. For the purpose of our rudimentary considerations, in the final stage of a hypothetical collapse, the Universe may be considered as a supermassive black hole. Deep inside the horizon of such a black hole, an extremely strong gravitational field can create particle-antiparticle pairs from the physical vacuum; with the additional feature that a black hole made from matter violently repels antiparticles, while a black hole made from antimatter repels particles. Without loss of generality we may consider the case of a black hole made from matter. The amount of created (and violently repelled) antimatter is equal to the decrease in the mass of black hole. Hence, during a Big Crunch, quantity of matter decreases while quantity of antimatter increases for the same amount; the final result might be conversion of nearly all matter into antimatter. If (as I will argue later) the process of conversion is very fast, it may look as a Big Bang starting with an initial size many orders of magnitude greater than the Planck length, which may be an alternative to the inflation in Cosmology.

The most poetic part of this qualitative picture is that a Big Crunch of a universe made from matter, leads to a Big-Bang-like birth of a new universe made from antimatter. Hence, the question why our Universe is dominated by matter has a simple and striking answer: because the previous universe was made from antimatter!

Let us consider the simplest case of a Schwarzschild black hole made from matter. While it is often neglected, from a mathematical point of view there are two solutions: the positive mass Schwarzschild solution

$$ds^2 = c^2\left(1 - \frac{2GM}{c^2 r}\right)dt^2 - \left(1 - \frac{2GM}{c^2 r}\right)^{-1} dr^2 - r^2 d\theta^2 - r^2 \sin^2\theta \, d\phi^2 \qquad (2)$$

considered as the physical space-time metric; and the negative mass Schwarzschild solution

$$ds^2 = c^2\left(1 + \frac{2GM}{c^2 r}\right)dt^2 - \left(1 + \frac{2GM}{c^2 r}\right)^{-1} dr^2 - r^2 d\theta^2 - r^2 \sin^2\theta \, d\phi^2 \qquad (3)$$

considered as a nonphysical solution. It serves as the simplest example of a naked singularity [16] and a repulsive space-time allowed by mathematical structure of general relativity but rejected as nonphysical. However, in the framework of the gravitational repulsion between matter and antimatter, both solutions may be given a physical meaning: the metric (2) is the metric "seen" by a test particle, while the metric (3) is the metric *"seen" by a test antiparticle*.

The major difference is that there is a horizon in the case of metric (2), while there is no horizon in the case of metric (3). In simple words, a black hole made from matter acts as a black hole with respect to matter and as a white hole with respect to antimatter.

According to the metric (3) the radial motion of a massive antiparticle is determined [14] by

$$\dot{r}^2 = c^2(k^2 - 1) - \frac{GM}{r} \qquad (4)$$

where *k* is a constant of motion and dot indicates the derivative with respect to the proper time.

Differentiating (4) with respect to proper time and dividing through by $\dot{r}$ gives

$$\ddot{r} = \frac{GM}{r^2} \qquad (5)$$

Equation (4) and (5) have the same form as should have the corresponding Newtonian equation of motion with the assumed gravitational repulsion; however, the Schwarzschild coordinate *r* is not identical with the radial distance in the Newtonian theory, and dots indicate derivatives with respect to proper time rather than universal time.

For simplicity, as a toy model [17], let's consider a black hole as a ball with decreasing "radius" $r_H < R_S \equiv 2GM/c^2$, and let us define a critical radius $r_{Cm} < R_S$, as the distance at which the gravitational acceleration $g = GM/r^2$, produced by a Schwarzschild black hole, has the critical value $a_{cr}(m) = \pi c^2 / \lambdabar_m$. Consequently,

$$r_{Cm} = \sqrt{\frac{\lambdabar_m R_S}{2\pi}} \qquad (6)$$

Hence a spherical shell with the inner radius $r_H$ and the outer radius $r_{Cm}$ acts as a "factory" for creation of particle-antiparticle pairs with mass *m*. It is evident that there is a series of decreasing critical radiuses $r_{Cm}$. For instance, according to (6), the critical radius $r_{Cv}$ corresponding to neutrinos is nearly four orders of magnitude greater than the critical radius $r_{Ce}$ for electrons, which is about 43 times greater than the critical radius $R_{Cn}$ for neutrons.

Integration of equation (1) over the shell determined by $r_H$ and $r_{Cm}$ (and taking $r_{Cm} >> r_H$) leads to the following approximation

$$\frac{dN_{m\bar{m}}}{dt} \approx \left(\frac{R_S}{\lambdabar_m}\right)^2 \frac{c}{r_H} \equiv \left(\frac{2GM}{c^2 \lambdabar_m}\right)^2 \frac{c}{r_H} \tag{7}$$

According to (7), the particle-antiparticle creation rate per unit time depends on both mass $M$ and radius $r_H$. If $r_H$ (i.e. the size of a black hole) is very small, the conversion of matter into antimatter is very fast!

As a numerical illustration let us calculate the number of created neutron-antineutron pairs in the case $M = 10^{54} kg$ and $r_H \approx 10^3 m$ (i.e. about $10^{38}$ Planck lengths!)

$$\frac{dN_{n\bar{n}}}{dt} \approx 10^{90} \, pairs/s \tag{8}$$

The numerical result (8) tells us that decrease of matter and increase of antimatter has a rate of about $10^{63} kg/s$, while the assumed mass of our Universe is "only" about $10^{54} kg$! Such a colossal conversion rate indicates that nearly the entire matter of the Universe might be transformed into antimatter (i.e. a Big Crunch of our Universe might be transformed to an event similar to Big Bang) in a tiny fraction of a second! According to this numerical example, the size of the newly born Universe should be about 38 orders of magnitude greater than the Planck length, suggesting that we do not need the inflation in Cosmology.

Let us give a second, presumably extreme but instrumental numerical example, taking $r_H \approx 10^{-15} m$, which is the size of a nucleon, but still $20$ orders of magnitude greater than the Planck length. Now, instead of (8), about $10^{108}$ neutron-antineutron pairs might be created per second, corresponding to an incredible conversion rate of $10^{81} kg/s$.

Hence, an eventual gravitational collapse of our Universe might end with the birth of a new Universe dominated by antimatter, with a macroscopic initial size, without inflation and the grand unification epoch.

Of course, this section would be incomplete without addressing the question if a future contraction of the Universe is really possible. According to the Standard Cosmological Model, the universe will continue to expand forever. However, it is an open question. A future collapse is predicted in different alternative cosmological models [18], including quantum loop cosmology [19] and cyclic models motivated by supersymmetries [20]. The important question is if a future reversal from expansion to contraction can be caused by the quantum vacuum enriched with gravitational dipoles. Apparently, the answer is positive, as we will argue in a much longer publication which will be finished within the next few months. The key point is that because of the existence of gravitational dipoles, the standard equations of state for dark matter and dark energy must be modified; one consequence of this modification is a future gravitational collapse.

## 3. Cosmological constant problem and virtual gravitational dipoles

The nature of dark energy, invoked to explain the accelerated expansion of the Universe, is a major mystery in theoretical physics and cosmology. From the purely mathematical point of view, adding a positive cosmological constant term to the right-hand side of the Einstein equation, can account for the observed accelerated expansion. However no one knows what is the physics behind such an ad hoc introduction of the cosmological constant. In principle, the cosmological constant $\Lambda$ may be interpreted as a cosmological fluid with a constant density $\rho_{de}$ and negative pressure ($p_{de} = -\rho_{de}c^2$) i.e. $\Lambda = 8\pi G \rho_{de}/c^2$ but the physical nature of such a hypothetical fluid remains unknown. The most elegant and natural solution would be to identify dark energy with the energy of the quantum vacuum predicted by Quantum Field Theory (QFT); but the trouble is that QFT predicts [9, 10] the energy density of the vacuum to be many orders of magnitude greater than the observed [21] dark energy density and the corresponding cosmological constant:

$$\rho_{de} \approx 7.1 \times 10^{-27} kg/m^3, \quad \rho_{de}c^2 \approx 6.4 \times 10^{-10} J/m^3 \tag{9}$$

$$\Lambda \approx 1.3 \times 10^{-52} m^{-2} \tag{10}$$

According to QFT, summing the zero-point energies of all normal modes [9, 10] of some field of mass $m$ up to a wave number cut-off $K_c \gg m$ yields a vacuum energy density (with $\hbar = c = 1$)

$$\langle \rho_{ve} \rangle = \int_0^{K_c} \frac{k^2 \sqrt{k^2 + m^2}}{(2\pi)^2} dk \approx \left(\frac{K_c}{2\pi}\right)^4 \tag{11}$$

or reintroducing $\hbar$ and $c$, and using the corresponding mass cut-off $M_c$ instead of $K_c$:

$$\rho_{ve} = \frac{1}{16\pi^2}\left(\frac{c}{\hbar}\right)^3 M_c^4 \equiv \frac{\pi}{2} \frac{M_c}{\lambda_{Mc}^3} \tag{12}$$

where $\lambda_{Mc}$ denotes the (non-reduced) Compton wavelength corresponding to $M_c$. If we take the Planck scale (i.e. the Planck mass) as a cut-off, the vacuum energy density calculated from (12) is $10^{121}$ times greater than the observed dark energy density (9). If we only worry about zero-point energies in quantum chromodynamics (i.e. if the cut-off mass is about the mass of a pion), (12) is still $10^{41}$ times larger than (9). Even if the Compton wavelength of an electron is taken as the cut-off, the result exceeds the observed value by nearly $30$ orders of magnitude. This huge discrepancy is known as the cosmological constant problem and it is the principal obstacle in the attempt to interpret dark energy as the energy of the quantum vacuum.

The result (12) is a completely wrong estimation of the gravitational charge density of the quantum vacuum, but, if we trust quantum field theory (and we have all reasons to trust it) it must be a correct estimation of the inertial mass density. Consequently, the incredible disagreement of the result (12) with observations can be considered as a strong hint that, for some unknown reasons, the inertial mass of the quantum vacuum is many orders of magnitude greater than the gravitational charge.

Now, let us assume that the gravitational charge $m_g$ of a particle and the gravitational charge $\bar{m}_g$ of an antiparticle have opposite sign (of course the corresponding inertial masses are equal $m_i = \bar{m}_i$). Consequently, a virtual particle-antiparticle pair in the quantum vacuum can be considered as a gravitational dipole (See also comments in Section 7), with the gravitational dipole moment

$$\vec{p}_g = m_g \vec{d}; \quad |\vec{p}_g| < \frac{\hbar}{c} \tag{13}$$

Here, by definition, the vector $\vec{d}$ is directed from the antiparticle to the particle, and has a magnitude $d$ equal to the distance between them. Consequently, a gravitational polarization density $\vec{P}_g$ (i.e. the gravitational dipole moment per unit volume) may be attributed to the quantum vacuum. The inequality in (13) follows from the fact that the size $d$ of the virtual pair must be smaller than the reduced Compton wavelength $\lambdabar_m = \hbar/mc$ (for a larger separation a virtual pair becomes real). Hence, $|\vec{p}_g|$ must be a fraction of $\hbar/c$.

The first fundamental consequence of the hypothesis (13) is: *Without matter immersed in it,* the gravitational charge density of the physical vacuum *is zero*. In fact, as we have already noticed, some virtual pairs in the quantum vacuum (like quark-antiquark and electron-positron pairs) are virtual electric dipoles. In the absence of an external electromagnetic field, these dipoles are randomly oriented and consequently the electric charge density of the quantum vacuum is zero. If the virtual pairs are gravitational dipoles (i.e. if particles and antiparticles, known to have the same inertial mass have the gravitational charge of the opposite sign) an analogous statement would also be true for gravitation: in the absence of external fields, the gravitational charge density (and consequently the cosmological constant) of the quantum vacuum is zero. This is the simplest candidate for the solution of the cosmological constant problem; without matter immersed in it, the quantum vacuum has a zero cosmological constant, while a small non-zero value emerges as a result of immersed matter.

## 4. The gravitational polarization density of the quantum vacuum

As already noted, without matter immersed in the quantum vacuum (i.e. without an external field), virtual dipoles are *randomly* oriented and the corresponding gravitational polarization density $\vec{P}_g$ is equal to zero. In an external gravitational field $\vec{g}$, the gravitational polarization density is *different* from zero: $\vec{P}_g \neq \vec{0}$.

While there is convincing evidence that the quantum vacuum exists, the current knowledge of its structure is very incomplete and does not permit development of a complete theory based on hypothesis (13). Fortunately, in spite of absence of detailed knowledge, we can make a few important conclusions.

As well known, in a dielectric medium the spatial variation of the electric polarization generates a charge density $\rho_b = -\nabla \cdot \vec{P}_e \equiv -div\vec{P}_e$, known as the bound charge density. In an analogous way, the gravitational polarization of the quantum vacuum should result in a *gravitational bound charge density* of the vacuum

$$\rho_{bg} = -\nabla \cdot \vec{P}_g \equiv -div\vec{P}_g \tag{14}$$

The potential energy of a gravitational dipole in an external gravitational field is equal to: $-\vec{p}_g \cdot \vec{g}$; hence the corresponding energy density is

$$\varepsilon_{gd} = -\vec{P}_g \cdot \vec{g} \tag{15}$$

The simplest possible case of the gravitational polarization of the quantum vacuum is *saturation* i.e. the case when the external gravitational field is sufficiently strong to align all dipoles along the field. If all dipoles are aligned in the same direction, the gravitational polarization density $\vec{P}_g$ has the maximal magnitude

$$\left|\vec{P}_g\right| \equiv P_{g\,max} = \frac{A}{\lambda_m^3}\frac{\hbar}{c} \tag{16}$$

where $A < 1$ should be a dimensionless constant of order of unity (as an approximation we adopt the value $A = 1/2\pi$ resulting from comparison with equations (19) and (20)). The relation (16) is a consequence of inequality (12) and the prediction of quantum field theory that the number density of the virtual gravitational dipoles has constant value $1/\lambda_m^3$.

It is more difficult to align more massive dipoles. Hence, for a given external field, dipoles with a sufficiently big mass will stay randomly oriented and will not contribute to the gravitational polarization density. In respect to the relations (11) and (12) it means that the cut-off value ($K_c$ or $M_c$) is not indeterminate (as naively considered in quantum field theory); the cut-off depends on the external field (i.e. on the distribution of matter immersed in the physical vacuum). In quantum field theory the cut-off is introduced to avoid an infinite value of the integral (11), while the hypothesis of virtual gravitational dipoles provides a *physical reason* for a cut-off and absence of infinity.

The mean distance between two dipoles which are the first neighbors is $\lambda_m$. The gravitational acceleration produced by a particle at the distance of its own Compton wavelength is

$$g_\lambda(m) = \frac{Gm}{\lambda_m^2} \qquad (17)$$

In absence of more accurate estimates, this acceleration can be used as a rough approximation of the external gravitational field which is needed to produce the effect of saturation for the dipoles of mass $m$. As an aside, the accelerations (17) corresponding to the Planck mass, a neutron and a pion are respectively: $5.7 \times 10^{51} \, m/s^2$, $6 \times 10^{-8} \, m/s^2$ and $2.1 \times 10^{-10} \, m/s^2$. For comparison: the acceleration corresponding to neutrons is about one order of magnitude greater than the current acceleration of the expansion of the Universe, while only in central parts of galaxies is the gravitational field stronger than acceleration corresponding to a pion. Hence, the acceleration corresponding to the Planck mass is about sixty orders of magnitude greater than typical gravitational fields in the present day Universe and cannot be the cut-off in (11); the relation (17) and the observed acceleration of the expansion of the Universe suggest that the right cut-off for the present day Universe should be close to the mass $m_\pi$ of a pion (which is a typical mass in the physical vacuum of quantum chromodynamics). In the following considerations, we will use as an approximation the mass $m_\pi$, while in a more accurate approach it would be necessary to consider quark and gluon condensates of quantum chromodynamics (with an effective mass slightly greater than $m_\pi$).

## 5. Dark energy and virtual gravitational dipoles

If the virtual gravitational dipoles exist, there are two intriguing ways to estimate the correct order of magnitude of the gravitational charge density of the quantum vacuum.

The first estimate is simply the result (12) multiplied by $\lambda_m/R_0$ where $R_o$ is the present day value, of the cosmic scale factor $R(t)$ in the Friedman–Lemaitre–Robertson–Walker metric.

$$\rho_{ve}^d \approx \frac{m_\pi}{\lambda_\pi^2 R_0} \qquad (18)$$

where we have used superscript $d$ to underline that (18) is the gravitational charge density of the physical vacuum corresponding to the hypothesis of gravitational dipoles.

It is easy to understand the motivation for this approximation. The result (12) is a consequence of the assumption that a virtual pair is composed from two *identical* gravitational monopoles, while according to hypothesis (13) a virtual pair is *composed* from two *different* gravitational monopoles having the gravitational charge of the *opposite sign*. The gravitational potential of a dipole, at a distance $\vec{r}$ from the center of the dipole is equal to the gravitational potential of a monopole multiplied by $\vec{d} \cdot \vec{r}_0/r$, where $\vec{r}_0$ is the unit vector (the calculation to demonstrate this is analogous to the well-known case of electric dipoles). In the case of a significant alignment of dipoles $\vec{d} \cdot \vec{r}_0 \approx d$, or, if we are interested only in the order of magnitude $\vec{d} \cdot \vec{r}_0 \approx \lambda_m$. The question remains what is the value of $r$. According to the cosmological principle, there are no privileged dipoles and $r$ must have a Universal value for all dipoles; a single universal distance that we have in disposition in FLRW metric is the cosmic scale factor $R(t)$.

The striking point is that the described correction of the result (12) really works; the relation (18) gives the correct order of magnitude, because the present day value of $R_0$ is [21] a few times $10^{27} m$

Unlike the result (18), the second estimate is independent from the quantum field calculations. Our universe is in a phase of accelerating expansion, with the present-day acceleration $(\ddot{R})_0$ which is determined by the cosmological field equations [14]; hence, the acceleration $(\ddot{R})_0$ should be used instead of $g$ in equation (15). Using $(\ddot{R})_0$ and combining equations (15) and (16) gives

$$\rho_{ve}^d c^2 = \frac{A}{\lambda_\pi^3} \frac{\hbar}{c} (\ddot{R})_0 \qquad (19)$$

which is once again a correct order of magnitude. There is an additional intriguing fact: if $A = 1/2\pi$, the relation (19) can be obtained from the Unruh [22, 23] temperature

$$k_B T = \frac{1}{2\pi} \frac{\hbar}{c} g \qquad (20)$$

dividing by $\lambda_\pi^3$ and using $g = (\ddot{R})_0$. Let us note that Unruh temperature is the temperature of the quantum vacuum measured by an accelerated observer moving with the acceleration $g$. Hence, *independently* of our hypothesis of virtual gravitational dipoles, what is called dark energy density in standard cosmology is numerically equal to Unruh temperature corresponding to $g = (\ddot{R})_0$ and divided by $\lambda_\pi^3$.

While we have written equations (16) and (19) only for the current value of the gravitational charge density, they strongly suggest that contrary to standard cosmology, what we call dark energy cannot be a constant, but must vary with the evolution of the Universe.

## 6. Dark matter and virtual gravitational dipoles

It is well established that gravitational field in a galaxy (and also in a cluster of galaxies) is much stronger than it should be according to our theory of gravity and the existing quantity of baryonic matter. According to mainstream opinion [21], the gravitational field in a galaxy is stronger because galaxies are immersed in halos of dark matter. If it exists, in order to fit the observations, dark matter within a halo must be distributed in a particular way: the quantity $M_{dm}(r)$ of dark matter within a sphere with a Galactocentric radius $r$ is nearly a linear function of $r$, i.e. the radial dark matter density $dM_{dm}(r)/dr$ is constant for a given galaxy.

Here, we will use equation (14) to give initial arguments that gravitational dipoles might explain the observed phenomena without invoking hypothetical dark matter. However, before we continue, let us point a fundamental difference between the hypothesis of dark matter and the hypothesis of the quantum vacuum filled with virtual gravitational dipoles.

In Standard Cosmology the quantity of dark matter in the Universe *is a constant* and the ratio of dark matter and baryonic matter is a constant as well. Consequently, on the cosmological scale, dark matter and baryonic matter are modeled with the same equation of state, as a pressureless fluid. But, if instead of dark matter, we have the gravitational polarization of the quantum vacuum by the immersed baryonic matter (we still can talk about an *effective*, but not real dark matter) the effects depend on the distribution of baryonic matter and the size of the Universe. Hence the gravitational polarization of the quantum vacuum cannot be mimicked with a constant quantity of the effective dark matter.

Now, for simplicity, let us consider an isolated spherical body, of baryonic mass $M_b$, immersed in the quantum vacuum [24, 25]. Assuming spherical symmetry, equation (14) reduces to

$$\rho_{bg} = \frac{1}{r^2} \frac{d}{dr}(r^2 P_g(r)), \quad P_g(r) \equiv |\vec{P}_g(r)| \geq 0 \qquad (21)$$

The gravitational polarization density $P_g(r)$ has maximal value $P_{g\,max}$ (in the region of saturation near the body determined with a characteristic radius $R_c$) and asymptotically approaches zero for large distances; between two limits ($P_{g\,max}$ and $0$) the function $P_g(r)$ decreases.

In the region of saturation ($r < R_c$), the equation (21) leads to

$$\rho_{bg} = \frac{2 P_{g\,max}}{r}, \quad r < R_c = \lambda_\pi \sqrt{\frac{\pi M_b}{m_\pi}} \tag{22}$$

The second of equations (22) is result of our previous work [24, 25] improved by the use of the equation (16) with $A = 1/2\pi$.

Our understanding of the quantum vacuum is not sufficient to find function $P_g(r)$ within the rigorous approach of quantum field theory. However, we may consider the gravitational polarization of the quantum vacuum as analogous to polarization of a dielectric in an external field, or a paramagnetic in an external magnetic field. If so, paramagnetic ideal gas, ideal gas of electric dipoles and ideal gas of gravitational dipoles are three mathematically equivalent models [26]. Consequently, the gravitational polarization density should be determined by the appropriate Brillouin function $B_J(x)$ as it is the case with magnetic and electric polarization density. The simplest Brillouin function is $\tanh(x)$, corresponding to the case $J = 1/2$. Hence, we may use approximation

$$P_g(r) = P_{g\,max} \tanh\left(\frac{R_c}{r}\right) \tag{23}$$

where $R_c$ is a characteristic radius. Equations (21) and (24) determine the quantity of *effective* dark matter within a sphere of radius $r$:

$$M_{dm}(r) = 4\pi r^2 P_{g\,max} \tanh\left(\frac{R_c}{r}\right) \tag{24}$$

with corresponding density distribution

$$\rho_{dm}(r) = \frac{P_{g\,max}}{r}\left[2\tanh\left(\frac{R_c}{r}\right) - \frac{R_c}{r}\frac{1}{\cosh^2\left(\frac{R_c}{r}\right)}\right] \tag{25}$$

In the absence of physical understanding of the phenomenon, the distribution of (real or *effective*?) dark matter in a galaxy is usually described by empirical laws (NFW profile, Einasto profile, Burkert profile…). Our work is in progress, but the preliminary results show that effective dark matter distribution given by equations (24) and (25) fits observational findings for different galaxies at least as well as the best existing empirical laws. Even if one day, the existence of virtual gravitational dipoles is dismissed by experiments, it remains striking that at least mathematically, the distribution of dark matter in a galaxy has such a similarity with an ideal gas of electric and magnetic dipoles.

The striking result is that, at distances greater than a characteristic radius $R_c$, the equation (26) reduces to

$$\rho_{dm}(r) = \frac{P_{g\,max} R_c}{r^2} \tag{26}$$

or, in other words, the gravitational polarization of the quantum vacuum produces effects that can be mimicked by an effective dark matter mass $M_{dm}(r)$, distributed with a constant radial mass density:

$$\rho_r \equiv \frac{dM_{dm}}{dr} = 4\pi P_{g\max} R_c = \frac{C}{\lambda_\pi}\sqrt{m_\pi M_b} \quad (27)$$

where $C$ is a dimensionless constant (in fact, the choice $A = 1/2\pi$ leads to $C = 1/\sqrt{\pi}$). Let us forget for the moment our hypothesis of virtual gravitational dipoles, independently of it there is a mysterious rule: Find the geometrical mean of mass of a pion ($m_\pi$) and baryonic mass ($M_b$) of a galaxy and divide it by the Compton wavelength ($\lambda_\pi = h/m_\pi c$) of a pion; what you get is very close to the value of the radial dark matter density! You can check it personally for every galaxy with measured dark matter distribution. In particular, for our Milky Way galaxy [27], with $M_b \approx 1.3 \times 10^{41} kg$, equation (24) gives $\rho_r \approx 3.8 \times 10^{21} kg/m^3$ and $M_{dm}(260kpc) \approx 3 \times 10^{42} kg = 1.5 \times 10^{12} M_{Sun}$ which is in surprising agreement with empirical evidence [27] $M_{dm}(260kpc) = 1 - 2 \times 10^{12} M_{Sun}$.

## 7. Discussion

We have given initial arguments that the physical vacuum enriched with virtual gravitational dipoles has the potential to explain a series of the most fundamental problems in physics, astrophysics and cosmology: What is the nature of what we call dark matter and dark energy? Why our Universe is dominated by matter? Why quantum field theory leads to the cosmological constant problem? If inflation existed or not in the primordial Universe? Was there an initial singularity?

Within this decade experiments at CERN [1-3] will reveal if antihydrogen has positive or negative gravitational charge, which would be a quantum leap in our understanding of gravity. No less important, in addition to laboratory experiments in the near future, the appropriate astronomical observations would be possible [28, 29].

So far, we have avoided the question how to assign the gravitational charges to ultimate constituents of matter. According to our best knowledge (i.e. the Standard Models of Particles and Fields), everything is composed from *three* generations of quarks and leptons. These fundamental building blocks are fermions (spin-½ particles) interacting through the exchange of gauge bosons (spin-1 particles): photons for electromagnetic interactions, gluons for strong interactions and $W^\pm$ and $Z^0$ bosons for weak interactions. Hence, the quantum vacuum should contain quark-antiquark and lepton-antilepton pairs, but also photons and other gauge bosons. However, we have no *experimental* answer on many questions, for instance, if the present day quantum vacuum contains quarks and leptons only from the first generation or from all three generations. Just to avoid any misunderstanding, let us say that gravitation is not the subject of the current Standard Model of Particles and Fields, it is simply neglected.

In the absence of any experimental evidence we can only speculate about gravitational charges of quarks, leptons and gauge bosons. Within the Standard Model there are six *quark-antiquark* pairs, six *lepton-antilepton* pairs, and, one pair of gauge bosons ($W^-$ and $W^+$) is also a particle-antiparticle pair. Obviously, under the assumption of the negative gravitational charge of antiparticles, all these pairs are gravitational dipoles. The situation with the neutral gauge bosons (photon, $Z^0$ and eight gluons) is not so evident; the simplest approach is to assume that they are also gravitational dipoles. This may seem a wild assumption, but, as a motivation for reflection, let us remember that in Quantum Chromodynamics, gluons are bicolour objects (i.e. a gluon carries both *colour* charge and *anti-colour* charge; for non-experts colour is short name for the charge that is the source of strong interactions). While we think that this is the most elegant and economic hypothesis, there are many different ways to assign gravitational charge to the fundamental building blocks of the Standard Model. For instance, negative gravitational charge may exist for all ultimate constituents excepting neutrinos (in this case equations (11) and (12) are correct for neutrinos and with neutrinos as the cut-off, these equations lead to the correct estimate of the cosmological constant).

Some other questions not addressed in the present paper will be considered in forthcoming publications. Here, we end with the following clarification. As already noted, it is more difficult to align more massive gravitational dipoles. Another significant category are electric dipoles which also persist in their random orientations. For these reasons, the dominant contribution to the gravitational charge of the physical vacuum of the present day Universe comes from electrically neutral gravitational dipoles.

We know that negative gravitational charge is widely considered as an unlikely outcome of the forthcoming experiments. However, imagination and simultaneous study of many different ideas are crucial for the progress of theoretical physics, astrophysics and cosmology. Keeping an open mind is especially important now, when the first three years of the LHC experiments at CERN have ended [30] with "the nightmare scenario"; all tests confirm the Standard Model of Particles so well that theorists have the nearly impossible task of looking for new physics without any available experimental guidance, and, with *supersymmetric* theories (a longtime dominant and privileged candidate for new physics) nearly excluded.

# Appendix: Theoretical debate on antimatter gravity and forthcoming experiments

In the present paper we have focussed on the study of *consequences* of the conjecture that the quantum vacuum contains virtual gravitational dipoles. Within the framework of our current understanding of the quantum vacuum, the simplest and the most elegant assumption is to attribute the hypothetical positive and negative gravitational charge respectively to virtual particles and antiparticles. However, some caution is needed; we still have to learn a lot about the content of the quantum vacuum and it is possible that the hypothesis of the existence of gravitational dipoles is more robust than the identification of dipoles with virtual particle-antiparticle pairs.

While the theoretical arguments against "antigravity" (an unfortunate name for the gravitational repulsion between matter and antimatter) are not topics of this paper, for completeness, in this Appendix we give a brief overview of theoretical debate concerning the gravitational properties of antimatter.

So far, arguments against antigravity are all based on three classical arguments suggested half a century ago (for a review see [31]). Morrison's argument [32] is a questionable attempt (in the form of a thought experiment) to show that antigravity is incompatible with the conservation of energy. Schiff's argument [33] is that because of the existence of the virtual particle-antiparticle pairs, different materials should contain *different fractions* of the virtual antimatter content; hence if antimatter falls up, it should be already detected by the classical tests of the weak equivalence principle. The third argument was developed by Good [34] who (before the discovery of CP violation) argued that antigravity must produce a very large CP violation. At the end of the last century, a critical reconsideration of the classical arguments [31] ended with the conclusion that these arguments are still sufficient to exclude antigravity, but also some serious shortcomings of the arguments were pointed out.

After the long domination of classical arguments against antigravity, the major turning point in the theoretical debate is the birth of *the first argument in the favour of antigravity* [35]: General Relativity and CPT theorem taken together lead to prediction of the gravitational repulsion between matter and antimatter.

Of course, only experiments and observations can tell us who is right. Hopefully, the answer will be known before the end of this decade. In addition to three experiments already approved at CERN [1-3], feasibility of some other experiments is under study. For instance, one outstanding proposal is to measure the gravitational acceleration of muonium [36]. The significance of muonium (an electron orbiting an antimuon) is in the opportunity to compare the gravitational properties of antileptons belonging to different generations (let us remember that quarks and leptons exist in three different generations).

While the experiments in our laboratories can reveal the gravitational properties of antimatter, only astronomical observations can establish if there is a gravitational impact of the quantum vacuum enriched with virtual gravitational dipoles; hence it is fortunate that in parallel with laboratory tests there are the first proposals [28,29] for astronomical tests within the Solar System.